\documentclass[%
 aip,
jcp,%
 amsmath,amssymb,
 reprint,%
 floatfix,
 10pt,twocolumn]{revtex4-1}
\usepackage{color}
\usepackage{graphicx}
\usepackage[caption=false]{subfig}
\usepackage{sidecap}
\usepackage{mathptmx}
\usepackage{algorithm}
\usepackage{natbib}
\usepackage{algorithmicx}
\usepackage[colorlinks=true,citecolor=blue,linkcolor=blue]{hyperref}%
\usepackage{epstopdf}
\usepackage{pstricks}
\usepackage{pst-plot}
\usepackage{pstricks-add}
\usepackage{epsf}

\expandafter\ifx\csname package@font\endcsname\relax\else
 \expandafter\expandafter
 \expandafter\usepackage
 \expandafter\expandafter
 \expandafter{\csname package@font\endcsname}%
\fi
\newcommand{\kB}{k_\mathrm{B}}

\makeatletter
\renewcommand{\fnum@algorithm}{\fname@algorithm}
\makeatother

\begin{document}
\title{Variational control forces for enhanced sampling of nonequilibrium molecular dynamics simulations}
\author{Avishek Das}
\affiliation{Department of Chemistry, University of California, Berkeley, California}
\author{David T. Limmer}
\affiliation{Department of Chemistry, University of California, Berkeley, California}
\affiliation{Kavli Energy NanoScience Institute, Berkeley, California}
\affiliation{Materials Science Division, Lawrence Berkeley National Laboratory, Berkeley, California}
\affiliation{Chemical Science Division, Lawrence Berkeley National Laboratory, Berkeley, California}
\date{\today}							
\begin{abstract}

We introduce a variational algorithm to estimate the likelihood of a rare event within a nonequilibrium molecular dynamics simulation through the evaluation of an optimal control force. Optimization of a control force within a chosen basis is made possible by explicit forms for the gradients of a cost function in terms of the susceptibility of driven trajectories to changes in variational parameters. We consider probabilities of time-integrated dynamical observables as characterized by their large deviation functions, and find that in many cases the variational estimate is quantitatively accurate. Additionally, we provide expressions to exactly correct the variational estimate that can be evaluated directly. We benchmark this algorithm against the numerically exact solution of a model of a driven particle in a periodic potential, where the control force can be represented with a complete basis. We then demonstrate the utility of the algorithm in a model of repulsive particles on a line, which undergo a dynamical phase transition, resulting in singular changes to the form of the optimal control force. In both systems, we find fast convergence and are able to evaluate large deviation functions with significant increases in statistical efficiency over alternative Monte Carlo approaches. 
\end{abstract}
\maketitle

\section{INTRODUCTION}

A system kept away from thermal equilibrium by a continuous supply of energy is subject to fewer physical constraints than one evolving within an equilibrium state. As a consequence, the application of external forces or the internal consumption of energy can produce structures and responses without equilibrium equivalent.\cite{nash2015topological,chakrabarti2017molecular,zwicker2017growth} 
Advances in the theory and modeling of nonequilibrium steady-states\cite{seifert2012stochastic,derrida2007non,prosen2009matrix} have resulted in an
 increased interest in trying to understand the behavior in systems out of equilibrium and leverage their versatility to design new functional materials.\cite{cates2015motility,rychkov2005block,bachelard2017emergence,abate2016shear,nguyen2016design,rotskoff2018robust,arango2018understanding} However, quantifying emergent nonequilibrium behavior with computer simulations is currently hampered by the lack of robust tools to sample the rare fluctuations required to estimate response functions, overcome kinetic bottlenecks, and reach the timescales of experimental relevance. For a generic class of stochastic systems that violate detailed balance, we have developed an algorithm to compute control forces that can be used to enhance the sampling of nonequilibrium steady-states. The control forces we optimize are variational, as are the estimates they provide of the likelihood of a rare fluctuation as characterized by large deviation functions.  In two paradigmatic models of nonequilibrium systems, we demonstrate that estimating large deviation functions in this way is both accurate and statistically efficient.

Enhanced sampling methods within equilibrium ensembles are standard tools that enable the determination of phase diagrams and the calculation of rates of rare events, through the evaluation of equilibrium free energies.\cite{frenkel2001understanding}
Free energies characterize the likelihood of configurational fluctuations around an equilibrium state, and the analogous quantity for fluctuations of time integrated observables around nonequilibrium steady-states are large deviation functions.\cite{touchette2009large,touchette2011basic,derrida2007non}  Large deviation functions have been used to map regions of stability for nonequilibrium phases,\cite{grandpre2018current,lips2018brownian} to elucidate complex dynamical behavior\cite{merolle2005space,chandler2010dynamics,laffargue2013large,kuptsov2011large,perez2019sampling} and infer nonlinear and multivariate response.\cite{gao2017transport,gao2019nonlinear,andrieux2004fluctuation,gaspard2013multivariate} Methods to compute large deviation functions in systems with many degrees of freedom have largely been restricted to Monte Carlo based approaches, including cloning,\cite{giardina2006direct,giardina2011simulating} Forward Flux Sampling,\cite{valeriani2007computing,allen2009forward}, nonequilibrium umbrella sampling,\cite{dickson2009nonequilibrium} list-based algorithms\cite{tchernookov2010list} and Transition Path Sampling.\cite{bolhuis2002transition,dellago1998efficient} Most current algorithms scale exponentially in computational effort the further the rare fluctuation is from the mean behavior, as apart from stratification or population dynamics, most do not employ additional importance sampling.\cite{dinner2018trajectory,ray2018importance,gingrich2015preserving,hidalgo2017finite}
Recent work adding control forces to importance sample trajectory based Monte Carlo has demonstrated that even an approximate force can greatly improve the efficiency of Monte Carlo methods in estimating large deviation functions.\cite{ray2018exact,nemoto2014computation,nemoto2016population}  
Consequently there has been much work to find approximate control forces analytically or through empirical arguments in both lattice-based and continuous systems\cite{jacobson2019direct,whitelam2019evolutionary,dolezal2019large} and several iterative effective force optimization techniques have been proposed with varying levels of generality or accuracy.\cite{hartmann2012efficient,quer2018automatic,nemoto2017finite,ferre2018adaptive} The control forces in general can have many-body components in interacting particle systems,\cite{hartel2018three,dolezal2019large} can be long-ranged in systems with dynamical phase transitions,\cite{simon2010asymmetric} and can stabilize otherwise metastable states.\cite{weber2014dynamical} 

For Markovian systems, there exists an optimal control force, which is the unique additional force having the smallest contribution to the path ensemble measure that can be added to the system to make a rare fluctuation typical.\cite{chetrite2015nonequilibrium,jack2015effective} This optimal control force satisfies several variational identities.\cite{chetrite2015variational} 
By deriving such a variational principle and explicit forms for the gradients required to optimize it, we develop an algorithm that approximates the control force sufficiently well so as to make quantitatively accurate estimates of the likelihood of rare events within nonequilibrium steady-states.
In this way, we generalize previous work on variational control of single particle systems to interacting, continuous force systems, bypassing the need for exponentially scaling Monte Carlo sampling. 
 Our algorithm is similar in strategy to the recent use of thermodynamic variational principles to compute equilibrium free energies,\cite{valsson2014variational} and to the Rayleigh-Ritz variational principle that others have used to nonperturbatively compute effective forces far from equilibrium.\cite{alexander1997rayleigh} The variational principle that underlies our algorithm is related to minimum-entropy production principles\cite{nemoto2011thermodynamic,nemoto2011variational} and the Donsker-Varadhan formula in Markov Stochastic processes.\cite{donsker1975asymptotic} While our variational estimate of the large deviation function is subject to errors associated with the representation of the control force, we derive exact corrections that can be evaluated straightforwardly. In the two systems studied, these corrections are easy to evaluate, as our control forces are sufficiently close to the optimal control forces to make these corrections perturbatively small. However, in cases where the corrections are large, we show that using optimized control forces in conjunction with standard Monte Carlo algorithms can increase the statistical efficiency in the estimatation of large deviation functions by orders of magnitude.  In this way, our algorithm is similar to the use of variationally optimized wavefunctions for quantum Diffusion Monte Carlo calculations.\cite{ceperley1986quantum}

\section{Enhanced sampling from optimal control forces}
Our aim is to construct a method by which rare fluctuations within a nonequilibrium steady-state can be sampled. 
We consider dynamics described by a Langevin equation of the form,
\begin{equation}\label{eq:EOM}
\dot{\mathbf{a}}=\mathbf{F}+\pmb{\eta}
\end{equation}
where \(\mathbf{a}\) is the vector of all dynamical coordinates, \(a_{i}\), which can include the positions and velocities of all particles in the system. Its time derivative, \(\dot{\mathbf{a}}\), depends on the force, \(\mathbf{F}\), with components, \(F_{i}\), that are in principle functions of all coordinates, \(\mathbf{a}\). The Gaussian white noise, \(\pmb{\eta}\), has components, \(\eta_{i}\), that satisfy
\begin{align}\label{eq:GWN}
&\langle\eta_{i}(t)\rangle=0 \,, \quad \langle\eta_{i}(t)\eta_{j}(t')\rangle=B_{i} \delta_{ij} \delta(t-t')
\end{align}
where \(B_{i}\) are diagonal elements of the diffusion constant matrix, \(\mathbf{B}\). While we have assumed \(\mathbf{B}\) is diagonal and independent of \(\mathbf{a}\) for ease of notation, generalizations for nondiagonal and coordinate-dependent diffusion matrices are straightfoward.

For a specific trajectory, \(X(\tau)=\{\mathbf{a}(0),...,\mathbf{a}(\tau)\}\) spanning an observation time, $\tau$, we are interested in fluctuations of time-averaged observables \(A_{\tau}\) of the form
\begin{equation}\label{eq:obs}
A_{\tau}[X(\tau)]=\frac{1}{\tau}\int_{0}^{\tau}dt\,f[\mathbf{a}(t)]+\frac{1}{\tau}\int_{0}^{\tau}dt\,\mathbf{g}[\mathbf{a}(t)]\cdot\mathbf{\dot{a}}(t)
\end{equation}
where \(f\) is a scalar function and \(\mathbf{g}\) is a vector function with components, \(g_{i}\), with the second term being evaluated in the Ito sense.\cite{rogers2000diffusions} Path observables like the particle density, particle current, and entropy production can all be expressed in this form. We will be interested in the statistics of this observable in the long time limit, $\tau \rightarrow \infty$.

\subsection{Nonequilibrium variational principle}
We assume that in the long time limit, the probability distribution 
of \(A_{\tau}\) satisfies a large deviation principle, with a rate function, or log likelihood, \(I(A)\), defined by\cite{touchette2009large}
\begin{equation}
I(A)=-\lim_{\tau\to\infty}\frac{1}{\tau}\ln\langle\delta(A-A_{\tau}[X(\tau)])\rangle
\end{equation}
where the angular brackets denote a trajectory average
\begin{equation}
\langle\delta(A-A_{\tau}[X(\tau)])\rangle=\int D[X(\tau)]\delta(A-A_{\tau}[X(\tau)])P[X(\tau)]
\end{equation}
and \(P[X(\tau)]\) denotes the path probability associated with trajectory \(X(\tau)\). We will consider finite size systems that have exponentially decaying correlation functions and thus are expected to obey the large deviation principle.  

The long time behavior of \(A_{\tau}\) can also be characterized by its scaled cumulant generating function (SCGF), defined as
\begin{equation}
\psi(\lambda)=\lim_{\tau\to\infty}\frac{1}{\tau}\ln\left\langle e^{\lambda\tau A_{\tau}}\right\rangle
\end{equation}
where \(\lambda\) is a counting parameter conjugate to \(A_{\tau}\), and denotes the extent of \textit{biasing} or \textit{tilting} on the typical value of  \(A_{\tau}\). Larger positive or negative values of \(\lambda\) probe rarer fluctuations. This is clear by noting that the derivatives of \(\psi(\lambda)\) report on the cumulants of \(A_{\tau}\). We refer to the rate function, \(I(A)\), and the SCGF,  \(\psi(\lambda)\), collectively as the large deviation functions. When the rate function is convex, it can be obtained from the SCGF using a Legendre-Fenchel transform
\begin{equation}
I(A)=\inf_{\lambda}[\lambda A-\psi(\lambda)]
\end{equation}
where \(\inf\) refers to an infimum taken over all possible values of \(\lambda\). 

Computing either of the large deviation functions of \(A_{\tau}\) requires sampling exponentially rare fluctuations. 
These rare fluctuations can in principle be made to occur more frequently by introducing a control force into the system as a means of importance sampling. In the presence of a new force, \(\mathbf{u}(\mathbf{a})\), replacing the original force, \(\mathbf{F}(\mathbf{a})\), the computation of the SCGF can done by changing the path ensemble measure,
\begin{align}\label{eq:cmlt}
\psi(\lambda)&=\lim_{\tau\to\infty}\frac{1}{\tau}\ln\int D[X(\tau)] e^{\lambda\tau A_{\tau}}\frac{P[X(\tau)]}{P_{\mathbf{u}}[X(\tau)]}P_{\mathbf{u}}[X(\tau)]\nonumber\\
&=\lim_{\tau\to\infty}\frac{1}{\tau}\ln\left\langle e^{O_{\tau}[\mathbf{u}]}\right\rangle _{\mathbf{u}}
\end{align}
where \(\langle\cdot\rangle_{\mathbf{u}}\) denotes an average in the controlled path ensemble with path probabilities \(P_{\mathbf{u}}[X(\tau)]\), and \(O_{\tau}[\mathbf{u}]\) can be derived from the difference in Onsager-Machlup path-actions,\cite{onsager1953fluctuations}
\begin{equation}\label{eq:cmlt2}
O_{\tau}[\mathbf{u}]=\lambda\tau A_{\tau}+\int_{0}^{\tau}dt\,\sum_{i}\frac{u_{i}^{2}-F_{i}^{2}-2\dot{a}_{i}(u_{i}-F_{i})}{2B_{i}}
\end{equation}
interpreted in the Ito sense. Changing the force for such a Gaussian process does not change the normalization constant associated with the path ensemble in the long time limit where boundary terms from the initial and final configurations can be ignored.

Expanding Eq. (\ref{eq:cmlt}) in terms of its cumulants, and using Jensen's inequality, we find a variational expression for the SCGF,
\begin{align}
\psi(\lambda)\geq \lim_{\tau\to\infty}\frac{1}{\tau}\langle O_{\tau}[\mathbf{u}]\rangle_{\mathbf{u}}
\end{align}
in terms of the mean of \(O_{\tau}[\mathbf{u}]\), within the controlled path ensemble. This expression is identical to previous work by Chetrite and Touchette that was derived using the contraction principle.\cite{chetrite2015variational}
Among the forces that make the rare value of the observable statistically typical, the one closest to the original force is the optimal force that realizes the supremum of the inequality. This many-body function can be approximated within a chosen ansatz with variationally optimizable parameters \(\{c_{n}\}\). In the limit that \(\{c_{n}\}\) represents all possible functional forms of the many-body force, this ansatz becomes exact,\cite{chetrite2015variational} so that
\begin{equation}\label{eq:sup}
\psi(\lambda)=\sup_{\{c_{1},c_{2},...\}}\lim_{\tau\to\infty}\frac{1}{\tau}\left\langle O_{\tau}[\mathbf{u}(\{c_{n}\})]\right\rangle _{\mathbf{u}(\{c_{n}\})}
\end{equation}
where the optimal coefficients \(\{c_{n}\}\) will in general depend on \(\lambda\). 

The existence of a control force that saturates the supremum in Eq.~(\ref{eq:sup}) follows from the eigenspectrum of the generator of the SCGF,
\begin{align}\label{eq:ltilt}
L_{\lambda}&=\lambda f+\sum_{i}\left[ -\frac{\lambda}{2}(\partial_{a_i}g_i)g_i +F_{i}(\partial_{a_i}+\lambda g_{i})\right. \nonumber\\
&\left. +\frac{B_{i}}{2}(\partial_{a_i}^{2}+\lambda(\partial_{a_{i}}g_i)+2\lambda g_i \partial_{a_i}+\lambda^{2} g_{i}^{2}) \right]
\end{align}
where we have suppressed the arguments of \(F_{i}\), \(g_{i}\), and \(f\) for compactness. This operator satisfies an eigenvalue equation
\begin{equation}
\label{eq:ltilt2}
L_{\lambda}\phi_{\lambda}(\mathbf{a})=\psi(\lambda)\phi_{\lambda}(\mathbf{a})
\end{equation}
where \(\psi(\lambda)\) and \(\phi_{\lambda}(\mathbf{a})\) are respectively the largest real eigenvalue and corresponding right eigenvector of \(L_{\lambda}\), which follows from the Perron-Frobenius theorem and the long time limit of the SCGF.
The optimal force \(\mathbf{u}_{\lambda}\) that solves Eq. (\ref{eq:sup}) is related to \(\phi_{\lambda}\) through a Hopf-Cole transform \cite{chetrite2013nonequilibrium,chetrite2015nonequilibrium,chetrite2015variational} defined as
\begin{equation}
\mathbf{u}_{\lambda}=\mathbf{F}+ \mathbf{B}(\lambda \mathbf{g}+\mathbf{\nabla}\ln\phi_{\lambda})
\end{equation}
and the controlled dynamics associated with this optimal force can be obtained from a generalized Doob transform of \(L_{\lambda}\).\cite{chetrite2015nonequilibrium,doob2012classical} For an interacting many-body system, the dominant eigenvector is a many body state, and therefore the optimal control force is many-bodied. Generally, we will assume that the control force is well approximated by a low rank ansatz such as obtained from a low order many body expansion.

Obtaining the SCGF from directly diagonalizing the tilted generator in many-body systems is prohibitively expensive due to the size of the multi-dimensional state space over which \(L_{\lambda}\) is defined. There have been recent advances to approximate this state space using  Matrix Product States for lattice based models. \cite{banuls2019using,helms2019dynamical} However, for continuous space systems with many particles, it is expected that Eq. (\ref{eq:sup}) will present a physically motivated way to formulate approximate solutions to the eigenvalue problem and to the computation of \(\psi(\lambda)\), and subsequently, \(I(A)\). It is worth noting that the constrained optimization of a variational expression analogous to (\ref{eq:sup}) can also be directly used to compute \(I(A)\),\cite{chetrite2015variational} with a straightforward extension of the algorithm described below.

\subsection{Optimization algorithm with explicit gradients}
In order to optimize Eq. (\ref{eq:sup}) by gradient descent, we need to calculate derivatives of \(\langle O_{\tau}[\mathbf{u}]\rangle_{\mathbf{u}}\) with respect to the variational parameters \(\{c_{n}\}\) in the limit of a large \(\tau\). 
Using these explicitly calculated gradients in the optimization algorithm can reduce the noise and numerical instabilities associated with finite difference schemes, that are generally used to empirically estimate the gradients from the optimization trajectory through the parameter space.
The explicit gradients that we use have the form of expectation values in the controlled ensemble,
\begin{align}\label{eq:derivative}
&\lim_{\tau\to\infty}\frac{1}{\tau}\frac{\partial}{\partial c_{n}}\left\langle O_{\tau}[\mathbf{u}]\right\rangle _{\mathbf{u}}\nonumber\\
&=\lim_{\tau\to\infty}\frac{1}{\tau}\left[ \left\langle \frac{\delta O_{\tau}[\mathbf{u}]}{\delta\mathbf{u}}\cdot\frac{\partial\mathbf{u}}{\partial c_{n}}\right\rangle _{\mathbf{u}}+\left\langle O_{\tau}[\mathbf{u}]\frac{\partial \ln P_{\mathbf{u}}}{\partial c_{n}} \right\rangle _{\mathbf{u}}\right]
\end{align}
where \(c_{n}\) is any of the optimizable parameters specifying the control force. While the first term is straightforward to compute, functional forms of \(\partial\ln P_{\mathbf{u}}/\partial c_{n}\) can be calculated from the normalized path probabilities,
\begin{align}\label{eq:odot}
&\left\langle O_{\tau}[\mathbf{u}]\frac{\partial \ln P_{\mathbf{u}}}{\partial c_{n}} \right\rangle _{\mathbf{u}}&\nonumber\\
&=\left\langle \int_{0}^{\tau} dt\,\dot{O}[\mathbf{u}](t)\int_{0}^{\tau} dt^{'}\,\sum_{i}\frac{\eta_{i}(t^{'})}{B_{i}}\frac{\partial u_{i}(t^{'})}{\partial c_{n}}  \right\rangle _{\mathbf{u}}\nonumber\\
&-\left\langle \int_{0}^{\tau} dt\,\dot{O}[\mathbf{u}](t)\right\rangle _{\mathbf{u}}\left\langle \int_{0}^{\tau} dt^{'}\,\sum_{i}\frac{\eta_{i}(t^{'})}{B_{i}}\frac{\partial u_{i}(t^{'})}{\partial c_{n}}  \right\rangle _{\mathbf{u}}
\end{align}
where Eqns. (\ref{eq:obs}) and (\ref{eq:cmlt2}) have been used to write \(O_{\tau}[\mathbf{u}]\) as a time integral of
\begin{equation}
\dot{O}_{\tau}[\mathbf{u}]= \lambda(f+\mathbf{g}\cdot\dot{\mathbf{a}})+\sum_{i}\frac{u_{i}^{2}-F_{i}^{2}-2\dot{a_{i}}(u_{i}-F_{i})}{2B_{i}}
\end{equation}
and its fluctuation is defined as $\delta \dot{O}_{\tau}[\mathbf{u}] = \dot{O}_{\tau}[\mathbf{u}] - \langle \dot{O}_{\tau}[\mathbf{u}] \rangle_{\mathbf{u}}$. The averages in Eq. (\ref{eq:odot}) can be computed by propagating additional coordinates \(y_{n}(t)\) associated with each variational parameter \(c_{n}\) as
\begin{align}\label{eq:malwt}
y_{n}(0)=0 \, ,\quad \dot{y}_{n}(t)=\sum_{i}\frac{\eta_{i}(t)}{B_{i}}\frac{\partial u_{i}(t)}{\partial c_{n}}
\end{align} 
where the sum has been performed over all dynamical coordinates of the system, and its fluctuation is defined as $\delta \dot{y}_{n}(t) = \dot{y}_{n}(t)- \langle \dot{y}_{n}(t)\rangle_{\mathbf{u}}$ .
These fictitious coordinates are known in the literature as Malliavin weights \cite{warren2014malliavin} and have previously been used to calculate parameter sensitivity of steady-state distributions in Langevin systems. \cite{warren2012malliavin} Provided these averages are evaluated in the steady-state generated by the control force, \(\dot{\mathbf{a}}=\mathbf{u}+\pmb{\eta}\), we can invoke time-translational invariance and note that only past noise history correlates with the observable, to simplify Eq.~(\ref{eq:derivative}),
\begin{align}\label{eq:tav1}
\lim_{\tau\to\infty}\frac{1}{\tau} \left\langle O_{\tau}[\mathbf{u}]\frac{\partial \ln P_{\mathbf{u}}}{\partial c_{n}} \right\rangle _{\mathbf{u}}&= \int_0^\infty dt \, \left\langle \delta \dot{y}_{n}(0) \delta \dot{O}[\mathbf{u}](t)\right\rangle _{\mathbf{u}}
\end{align}
where in the long time limit, the gradient is proportional to an integrated time correlation function. This is an example of a generalized fluctuation-dissipation formula.\cite{kubo2012statistical}
Putting together the two contributions 
\begin{align}\label{eq:tav2}
\lim_{\tau\to\infty}\frac{1}{\tau} \frac{\partial}{\partial c_{n}}\left\langle O_{\tau}[\mathbf{u}]\right\rangle _{\mathbf{u}}&=\\
\left\langle \frac{\delta \dot{O}_{\tau}[\mathbf{u}]}{\delta\mathbf{u}}\cdot\frac{\partial\mathbf{u}}{\partial c_{n}}\right\rangle _{\mathbf{u}} &+ \int_0^{\infty} dt \, \left\langle \delta \dot{y}_{n}(0) \delta\dot{O}[\mathbf{u}](t)\right\rangle _{\mathbf{u}} \nonumber
\end{align}
we arrive at an explicit form for the gradient of our SCGF estimate with respect to the variational parameters that can be estimated as time-averages from a straightforward molecular dynamics trajectory with the control forces. In practice, we will take the integral over the time correlation function in Eq.~(\ref{eq:tav2}) up to a time $\Delta t$. The choice of $\Delta t$ is discussed in Appendix A.

Using these explicit gradients, an iterative optimization is performed in the parameter space spanned by \(\{c_{n}\}\) in order to estimate the SCGF. We use an algorithm called
 Nesterov's Accelerated Gradient Descent \cite{nesterov1983method,sutskever2013importance} which shows a superlinear convergence. The learning rate and conjugate momenta are scaled by fixed parameters \(\mu\) and \(\nu\) respectively. The optimization algorithm is summarized below.
\begin{algorithm}[H]
  \caption{Optimizing control force}
  \label{alg}
   \begin{algorithmic}[1]
   \State Begin from a guess for the variational parameters \(\{c_{n}\}\) and conjugate momenta \(\{p_{n}=0\}\).
   \State After the \(\textit{k}\)-th step of the optimization, parametrize the force \(\mathbf{u}^{(k)}\) with parameters \(\{c^{(k)}_{n}+\nu p_{n}^{(k)}\}\).
   \State Propagate an MD trajectory to evaluate the gradients \(d_{n}^{(k)}=\partial[\left\langle O_{\tau}[\mathbf{u}]\right\rangle _{\mathbf{u}^{(k)}}/\tau]/\partial c_{n}\) for a large \(\tau\).
	\State Update the momenta as \(p_{n}^{k+1}=\nu p_{n}^{(k)}+\mu d_{n}^{(k)}\). 
	\State Update the variational parameters as \(c^{(k+1)}_{n}=c^{(k)}_{n}+p_{n}^{(k+1)}\).
	\State Repeat steps (2-5) until all \( \left |d_{n}^{(k)} \right |\) are less than a tolerance value.
   \end{algorithmic}
\end{algorithm}
This algorithm converges to a local maximum in the parameter space, which can be different from the global maximum when the variational surface is not convex. For all the models for which we computed the SCGF, we did not obtain evidence of nonconvexity of the variational functional at any point in the parameter space. However the convergence was significantly slower at values of \(\lambda\) near a crossover point or a phase transition. We have illustrated in Appendix B that we often converge to the global maximum in the parameter space smoothly. Nevertheless, in the event that we converge to a local maximum, we incur a systematic error in the SCGF that we discuss how to correct in the next section. 

\subsection{Correcting for systematic errors}

In general, the ansatz specified by the parameters \(\{c_{n}\}\) will not form a complete basis for a many body system. This is because generically, the dominant eigenvector of Eq. (\ref{eq:ltilt2}) is a many-body state, containing exponentially many parameters, and not expected to be exactly expressible with a low rank form.  Because of this, the variationally converged SCGF \(\psi^{*}(\lambda)\) obtained from Eq. (\ref{eq:sup}) will have a systematic error.  This error, and errors associated with convergence to a local maximum, can both be corrected in principle by computing the remaining terms of the cumulant expansion
\begin{equation}\label{eq:correction}
\psi(\lambda)=\psi^{*}(\lambda)+\lim_{\tau\to\infty}\frac{1}{\tau}\sum_{\ell=2}^{\infty}\frac{\kappa_{\ell}}{\ell!}
\end{equation}
where \(\{ \kappa_{\ell} \} \) are the second and higher cumulants in the expansion of \(\ln\langle \exp(O_{\tau}[\mathbf{u}^{*}])\rangle_{\mathbf{u}^{*}}\) and the force \(\mathbf{u}^{*}\) is the solution of the variational problem in the approximate and incomplete ansatz.
If the ansatz used to express the control force, \(\mathbf{u}^{*}\), is close enough to the optimal force obtained from the Doob transform, the correction terms are small in magnitude and the series will converge quickly. This will occur when the trajectory distribution generated by the controlled dynamics has significant overlap with the tilted distribution of the original dynamics. 

In cases where the ansatz is poor and many cumulants are needed, brute force convergence of the correction will be difficult. In such cases, control forces can be used as guiding functions for estimating the SCGF through Monte Carlo based approaches like the cloning algorithm. In the cloning algorithm, an ensemble of \(N_{w}\) trajectories generated from the ordinary path probabilities \(P[X(\tau)]\) are branched with corresponding weights of \(\exp(\lambda\tau A_{\tau})\). However, under the controlled dynamics, following Eq. (\ref{eq:cmlt}), the weighted path probabilities can be written as \cite{ray2018exact,dolezal2019large}
\begin{equation}
P_{\lambda}[X(\tau)]\propto e^{\lambda\tau A_{\tau}}P[X(\tau)]=e^{O_{\tau}[\mathbf{u}]}P_{\mathbf{u}}[X(\tau)]
\end{equation}
where system evolution under an approximate controlled dynamics is nonconservative and must be accompanied by branching steps with weights given by \(\exp( O_{\tau})\).
An estimate of the SCGF is then obtained from the normalization constant of this weight, so that in the limit of large \(N_{w}\),
\begin{equation}
\psi(\lambda)=\frac{1}{\tau}\ln\frac{1}{N_{w}}\sum_{j=1}^{N_{w}}e^{O_{\tau}^{(j)}[\mathbf{u}]}
\end{equation}
where \(O_{\tau}^{(j)}[\mathbf{u}]\) denotes the time-integrated observable for the walker labelled as \(j\).

When the variationally optimized \(\mathbf{u}^{*}\)  is used to generate trajectories and to compute the branching probabilities, the efficiency of the cloning algorithm is improved as the control force samples the rare fluctuations in the observable. When \(\mathbf{u}^{*}\)  is actually the optimal force derived from the Doob transform, all trajectories achieve the rare fluctuation as typical behavior, and the weight of each trajectory becomes a constant. In this situation no trajectories are killed in the branching step of the cloning algorithm, and the sampling is statistically optimal.\cite{giardina2006direct} However, even with an approximate ansatz the variationally optimized force slows down the rate of death of uncorrelated trajectories with increasing \(\tau\), as demonstrated in Sec. IIIB .

The variational algorithm along with the cumulant-correction has improved scaling properties compared to the cloning algorithm. By adopting an approximate ansatz for the many-body force containing, for example, one-body and two-body terms, for a system of identical particles we can exploit their permutation symmetry and optimize a single one-body and two-body force. Hence the variational algorithm scales linearly with the system size, the computational cost arising only from the propagation of trajectories of interacting particles. This is in contrast to the cloning algorithm, which has an exponential scaling for observables that are system size extensive.\cite{ray2018importance} Also, while the cloning algorithm scales exponentially with \(\lambda\), the variational algorithm depends on the bias only through the complexity of the optimal force and scales linearly with the number of variational parameters required to approximate the force. Hence in cases that the dominant part of the optimal force can be simply expressed within the choice of the ansatz, the computational cost for the algorithm to converge does not increase with \(\lambda\). This indicates a resummation of the exponential bias through the modification of the control force. Neither does the algorithm scale with increasing observation time \(\tau\), as the \(\tau\to\infty\) limit has already been incorporated in the algorithm. Lastly, this algorithm can be parallelized trivially by distributing the computation of the expectation values at each step of the iteration to independent trajectories on independent processors.

\section{Numerical Illustrations}
To study the accuracy and efficiency of our variational algorithm to compute the SCGF and the optimal force, we apply it to two different continuous time and space systems. The first is a benchmark system where we can test our algorithm against a numerically exact result. This model consists of a driven underdamped particle in a periodic potential, for which we have studied rare fluctuations of the total current. The second system is comprised of multiple repulsive overdamped particles, where we have focused on the fluctuations of the total activity, which measures how much the particles explore configuration space. In this system, we demonstrate the ability of our algorithm to compute the optimal control force even through singular changes in the SCGF across a dynamical phase transition. 

\subsection{Driven underdamped particle in a periodic potential}
An underdamped particle being driven on a periodic potential by a constant external force is a simple system with two dynamical coordinates, position and velocity, that can exhibit non-trivial nonequilibrium properties due to competing ballistic and diffusive modes of transport.\cite{shinagawa2016enhanced,ma2017colloidal} Large deviation functions for current fluctuations in this model can be obtained by numerically exact diagonalizations of the tilted generator, and the controlled ensemble can show diverse behavior in different parameter regimes.\cite{fischer2018large} We consider this model to benchmark our variational optimization algorithm.

Specifically, we consider an underdamped particle of mass \(m\) moving in a one-dimenional periodic box of length \(L=2\pi\). The forces acting on the particle are derived from a cosine potential, \(V(x)=V_{0}\cos (x)\), where \(V_{0}\) is the magnitude of the potential, and include a constant external driving force, \(F_{\mathrm{ext}}\). For the particle in contact with a bath of temperature, \(T\), and friction coefficient, \(\gamma\), the equations of motion for the position, \(x\), and velocity, \(v\), are
\begin{align}\label{eq:l1}
\dot{x}&=v\nonumber \\
m\dot{v}&=F(x)-\gamma v+\eta
\end{align}
where $F(x)=-V^{'}(x)+F_{\mathrm{ext}}$ and \(\eta(t)\) is a Gaussian white noise with 
\begin{equation}
\langle\eta(t)\rangle=0\hspace{1cm}\langle\eta(t)\eta(t')\rangle=2\gamma \kB T\delta(t-t')
\end{equation}
where $\kB$ is Boltzmann's constant. These equations of motion have the form of Eqs. (\ref{eq:EOM}) and (\ref{eq:GWN}) with two dynamical coordinates and a vanishing noise in position.\cite{fischer2018large}

We investigate the statistics of the time-averaged current flowing through the system,
\begin{equation}\label{eq:j1}
J_{\tau}=\frac{1}{\tau}\int_{0}^{\tau}dt\,v(t) 
\end{equation}
which measures the total displacement of the particle.
The SCGF for current is given by 
\begin{align}\label{eq:j2}
\psi(\lambda)=\sup_{u(x,v)}\frac{1}{\tau}\biggl\langle \int_{0}^{\tau}dt\biggl( \lambda v &+\frac{u^{2}-F^{2}-2\gamma v(u-F)}{4\gamma \kB T} \nonumber \\
&-\frac{m\dot{v}(u-F)}{2\gamma \kB T} \biggr) \biggr\rangle _{u}
\end{align}
where the path average is obtained from the controlled dynamics
\begin{equation}
m\dot{v}=u(x,v)-\gamma v+\eta 
\end{equation}
and the optimal force is in general a function of both position and velocity. We expand this force in an ansatz
\begin{equation}
\tilde{u}(x,v)=F(x)+\sum_{p=-M_{1}}^{M_{1}}\sum_{q=0}^{M_{2}} c_{p,q} e^{ipx}v^{q}
\end{equation}
where \(c_{pq}\) are parameters that can be optimized variationally subject to \(c^{*}_{-p,q}=c_{p,q}\), and the number of position and velocity basis functions are \((2M_{1}+1)\) and \((M_{2}+1)\) respectively. The basis is complete in the limit of large \(M_{1}\) and \(M_{2}\). Note that this force incorporates the periodicity of $x$ and also allows the external nonequilibrium driving, which is the \(p=q=0\) term, to be optimized. In the high friction limit, the dynamics becomes overdamped and in that limit the optimal force becomes a function of just the particle position. For small friction, inertia is important and the general form of the optimal force must be considered. We note that this velocity-dependent drift function is a \emph{force} only in a generalized sense.

The SCGF and the optimized control force obtained from the variational algorithm can be compared to numerically exact results obtained by solving the eigenvalue equation for the tilted generator given by \cite{fischer2018large}
\begin{equation}\label{eq:currentltilt}
L_{\lambda}=v\frac{\partial }{\partial x}-\frac{1}{m}[v\gamma-F(x)]\frac{\partial}{\partial v}+\frac{\gamma\kB T}{m^{2}}\frac{\partial^{2}}{\partial v^{2}}+\lambda v 
\end{equation}
as in Eq.~(\ref{eq:ltilt2}). The exact control force is obtained using the right eigenvector \(\phi_{\lambda}(x,v)\) corresponding to the largest real eigenvalue, as
\begin{equation}
u(x,v)=F(x)+\frac{2\gamma \kB T}{m}\frac{\partial \ln \phi_{\lambda}(x,v)}{\partial v}
\end{equation}
where numerical diagonalization of \(L_{\lambda}\) can be performed by expressing the right and left eigenvectors over a position-velocity grid and representing the differential operators in \(L_{\lambda}\) using a second order finite difference scheme. The boundary conditions are periodic in the position grid and reflective in the velocity grid, so that only forward (backward) difference at the minimum (maximum) velocity grid point is used to represent the differential operator.

We have computed the cumulant-corrected large deviation functions in this system and have compared them to the numerically exact results. We have worked with \(\kB T=1\) and \(\gamma=1\). These parameters along with the length of the box \(L=2\pi\) let us define our natural time unit as \(t^{*}=4 \pi^{2}\kB T/\gamma L^{2}\). All observables have been reported in dimensionless units following these definitions. We have done our computations at two values of mass, \(m/\gamma t^{*}=1\) and \(m/\gamma t^{*}\to0\). We have also chosen \(V_{0}=2\) and \(F_{\mathrm{ext}}=1\). The numerically exact result was obtained with a grid of \(140\times50\) points in the position-velocity space. The position points span all of the box and the velocity points are centered at \((F_{\mathrm{ext}}+2\lambda \kB T)/\gamma\) corresponding to the mean velocity in the \(V_{0}\to 0\) limit. For all the simulations, the timestep was chosen to be \(0.001\) natural time units. For \(m/\gamma t^{*}\to0\), an Euler scheme was used to integrate the overdamped equation of motion, while for \(m/\gamma t^{*}=1\), a velocity Verlet scheme was used.\cite{frenkel2001understanding}

\begin{figure}[t]
\includegraphics[scale=0.3]{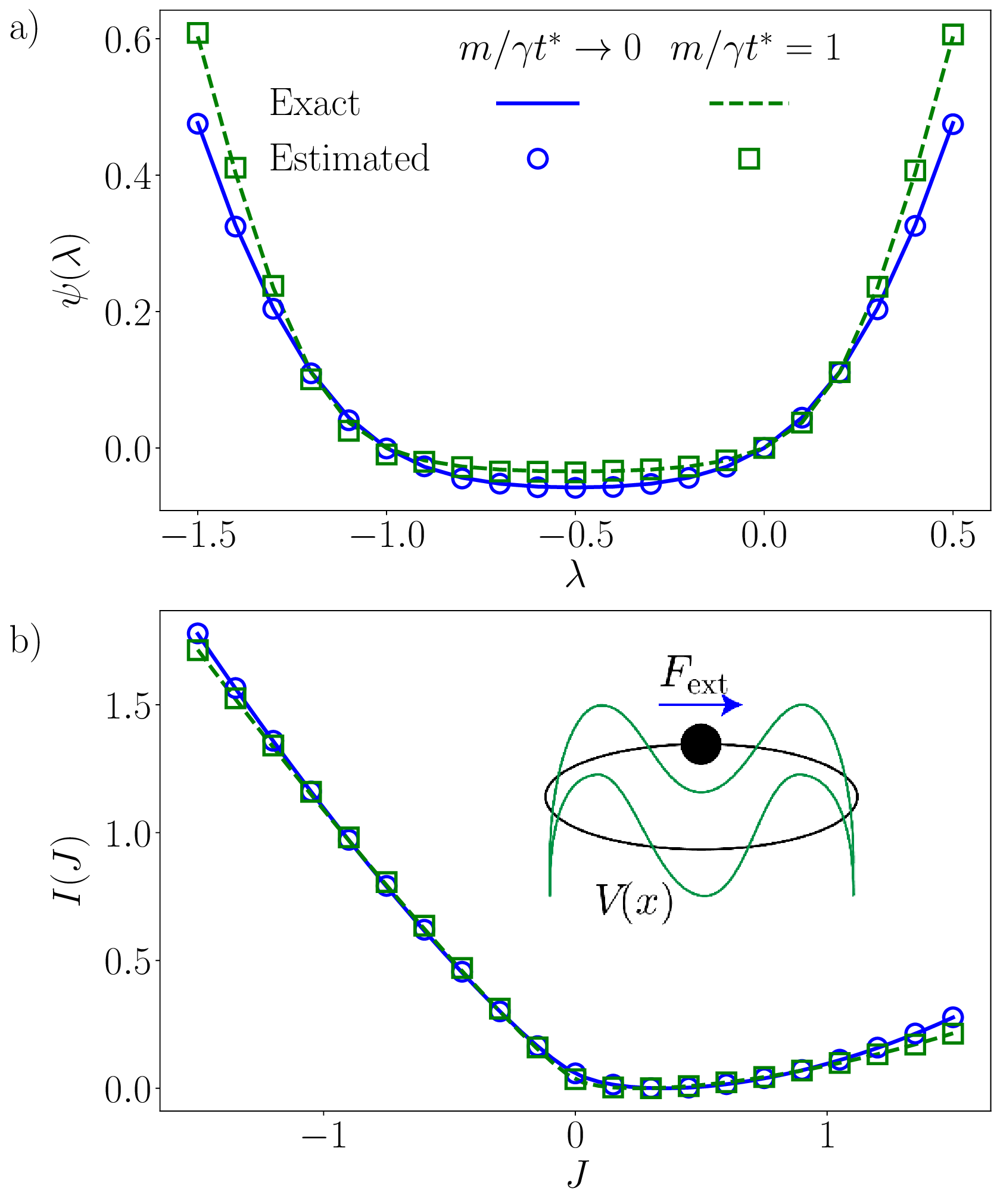}
\caption{Large deviation functions for current fluctuations in a driven underdamped system in a periodic potential. a) SCGF for \(m/\gamma t^{*}=1\) with \(M_{1}=3,M_{2}=1\) and for \(m/\gamma t^{*}\to0\) with \(M_{1}=3,M_{2}=0\). b) Rate functions obtained by a numerical Legendre-Fenchel transform of the SCGFs. The legend is the same as that used in a). (\textit{Inset}) Schematic diagram of the simulated system. }
\label{Fig1}
\end{figure}

For each iterative step during the optimization, a trajectory of duration \(10^{4}\) units was simulated. 
During the first half of each trajectory, the system was allowed to come to a steady-state, and the time-averaged gradients were computed only with the second half of the trajectories. For implementing Eq. (\ref{eq:tav2}), we integrated the correlation function up to \(\Delta t=100\). The size of the basis was \(M_{1}=3,M_{2}=1\) for \(m/\gamma t^{*}=1\) and \(M_{1}=3,M_{2}=0\) for \(m/\gamma t^{*}\to0\), the overdamped limit. The optimization parameters used for the gradient descent were \(\mu=0.5,\nu=0.2\). 
Near \(\lambda=0\), all \(c_{pq}\) were initialized at zero, and subsequent optimizations with increasing magnitude of \(\lambda\) were initialized from a previously optimized set of \(c_{pq}\) taken from the nearest value of \(\lambda\).
 In the overdamped limit, an accurate estimate of the SCGF could be obtained with just the variational optimization, with the cumulant correction merely a confirmation of the optimal control forces being correct. However for \(m/\gamma t^{*}=1\), the variational SCGF had to be corrected with cumulants computed with an observation time \(\tau=100\) and a total trajectory length \(10^{5}\) units. Following this procedure, we obtain estimates of SCGFs that are in quantitative agreement with the numerically exact results throughout the range of \(\lambda\) considered, as shown in Fig.~\ref{Fig1}(a). We have also calculated the rate functions for the current, Fig. ~\ref{Fig1}(b), in these two parameter regimes by a numerical Legendre-Fenchel transform of the SCGFs.

\begin{figure}[t]
\includegraphics[scale=0.3]{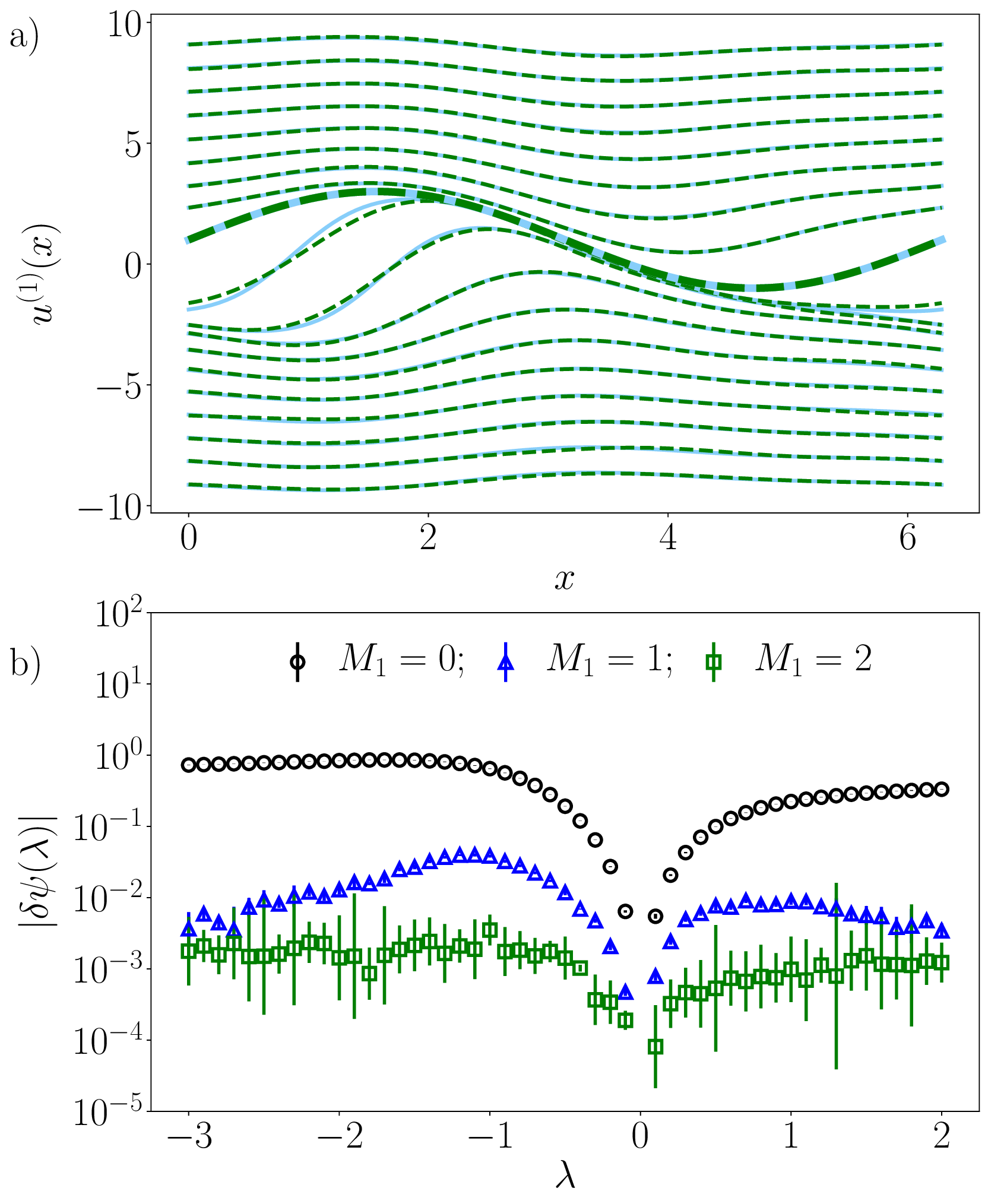}
\caption{Overdamped limit, \(m/\gamma t^{*}\to 0 \), of the driven particle on a periodic potential. a) Optimized control forces (dashed lines) overlaid on the exact control force (solid lines). The thick curve is for \(\lambda=0\) and the curves above (below) are for \(\lambda\) in intervals of \(+0.5\)(\(-0.5\)).  b) Basis size errors in the variational estimate of $\psi(\lambda)$, where the deviation $\delta \psi(\lambda)=\psi^{*}(\lambda)-\psi(\lambda)$ is the difference between the finite basis result $\psi^{*}(\lambda)$ from the exact SCGF.}
\label{Fig2}
\end{figure}

The SCGFs in Fig.~\ref{Fig1}(a) both have a \emph{locked} region where the current changes slowly with \(\lambda\), and an \emph{unlocked} region for larger magnitudes of \(\lambda\). Due to the time-reversal properties of \(L_{\lambda}\), the SCGF shows a Gallavotti-Cohen symmetry\cite{kurchan1998fluctuation}
\begin{equation}
\psi(\lambda)=\psi(-F_{\mathrm{ext}}/\kB T-\lambda)
\end{equation}
which is clear through the reflection symmetry about \(\lambda=-0.5\) of the SCGF in Fig.~\ref{Fig1}(a). Analogously, the rate function obeys a fluctuation theorem symmetry 
\begin{equation}
I(J)= I(-J) +F_{\mathrm{ext}}J/\kB T 
\end{equation}
indicating the exponentially rare probability of a current in the direction opposite to the applied force. 
\begin{figure}[b]
\includegraphics[scale=0.3]{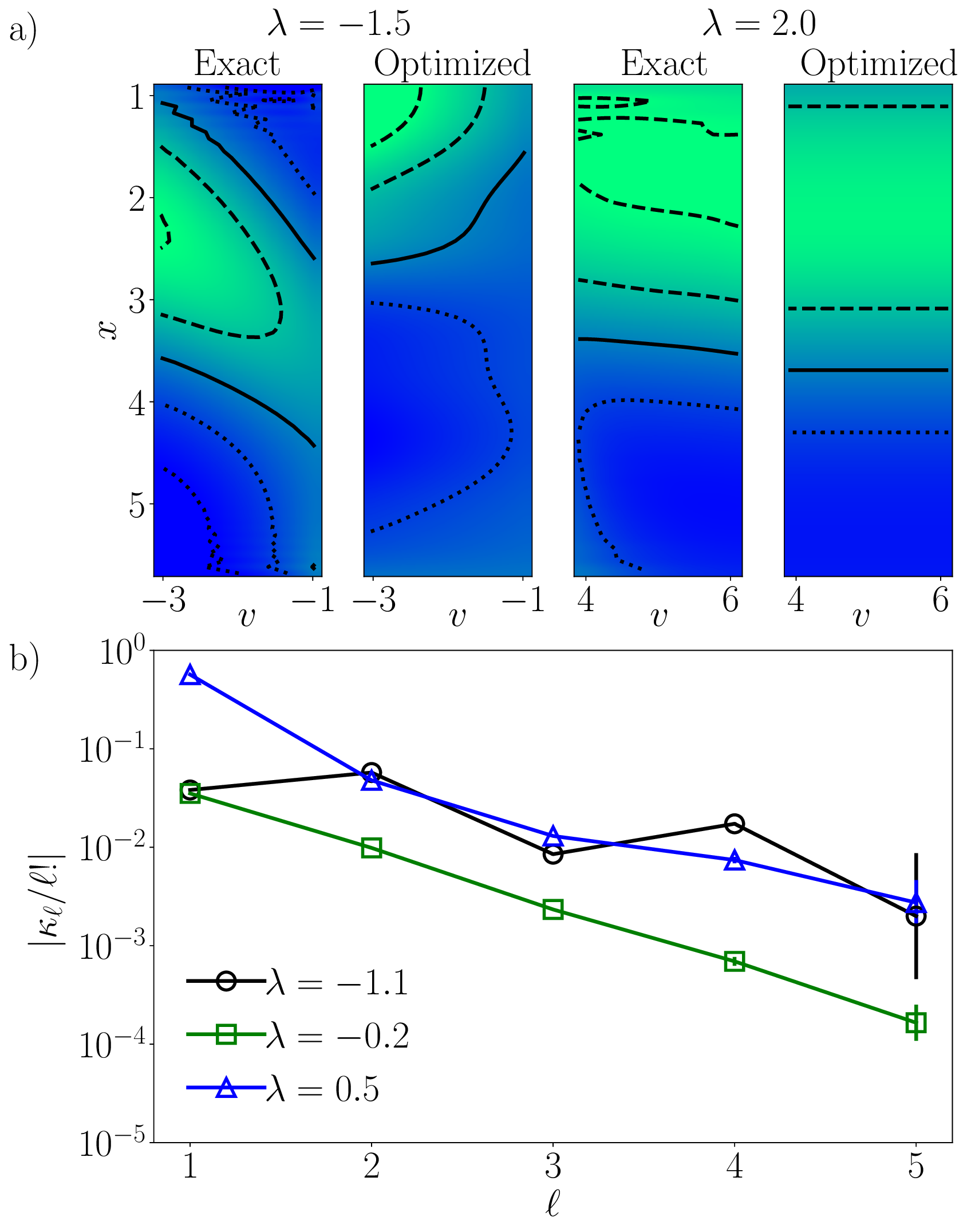}
\caption{Underdamped system, \(m/\gamma t^{*}=1\), of the driven particle on a periodic potential. a) (L-R) Exact and optimized control forces, $u(x,v)$, for \(\lambda=-1.5\), with the solid contour at \(u(x,v)=-2\), and the dashed (dotted) contours being at differences of \(+1\) (\(-1\)).  Exact and optimized control forces, $u(x,v)$, for \(\lambda=2\) with the solid contour at \(u(x,v)=5\) and the dashed (dotted) contours being at differences of \(+1\) (\(-1\)). b) Convergence of the cumulant expansion for representative values of \(\lambda\).}
\label{Fig3}
\end{figure}

Figure~\ref{Fig2}(a) shows the position-dependent optimal forces obtained in the overdamped limit, $u^{(1)}(x)$, overlaid on the numerically exact answers obtained from diagonalization,\cite{nyawo2016large} for multiple values of \(\lambda\). In the limit of \(|\lambda|\to\infty\), the optimal forces approach the free-diffusion limit, where the majority of contribution comes from a constant nonequilibrium driving. When \(|\lambda|\) is of the order \(|F_{\mathrm{ext}}|/\kB T\), the forces have a non-trivial position dependence. This is manifested in the size of the basis-set, \(M_{1}\), required to obtain the optimal control force accurately. Figure ~\ref{Fig2}(b) shows the effect of finite basis size on the error made in estimating \(\psi(\lambda)\). Increasing \(M_{1}\) reduces the error and ultimately the ansatz becomes exact when \(M_{1}\) is large. The error decreases when going to larger \(|\lambda|\) as the forces are easier to represent using the first few basis functions. The error bars were computed from 5 independent estimates of the SCGF using independent trajectories.

For the \(m/\gamma t^{*}=1\) system, inertial effects are important and the optimal force depends on both position and velocity, and the optimal force has a complicated functional dependency that is difficult to represent using a small number of basis functions. Using a truncated basis to represent the control force leads to a systematic error in the SCGF estimate obtained using Eq. (\ref{eq:j2}) that can be corrected using the cumulant expansion in Eq.~(\ref{eq:correction}).
Figure~\ref{Fig3}(a) shows the approximate forces obtained from the variational optimization compared to the numerically exact results. When \(\lambda\) is near the Gallavotti-Cohen symmetry point, the average current is small and the optimal control force is a complicated function of both $v$ and $x$. Within our ansatz, the optimized \(u(x,v)\) does not reproduce the exact form of the optimal control force. Nevertheless, these approximate forces recover the majority of the SCGF, so that the cumulant expansion converges for all tested \(\lambda\) points. Figure~\ref{Fig3}(a) also contains the optimal force at a larger positive \(\lambda\), where the forces lose their velocity dependence and simplify towards the free-diffusion limit. In this limit, position based forces are sufficient to recover the SCGF quantitatively.

Figure~\ref{Fig3}(b) shows the convergence of the consecutive terms of the cumulant expansion in Eq. (\ref{eq:correction}) for different values of \(\lambda\). \(\kappa_{1}\), the first cumulant, is identical to \(\psi^{*}(\lambda)\), the variational estimate. Error bars were calculated using 5 independent trajectories for the estimation of the cumulants. Even though our basis is small and approximate, the cumulants computed from a single trajectory have decreasing amplitudes for various values \(\lambda\), showing that the variational force is accurate enough to approach the force derived from the Doob transform. We note that the sign of the cumulants need not be positive, and therefore the variational structure in the estimate of $\psi(\lambda)$ holds only for the first cumulant. 
Further, the magnitude of the terms in the cumulant expansion need not be strictly decreasing. Figure~\ref{Fig3}(b) includes an example of a nonmonotonic convergence for \(\lambda=-1.1\).
Moreover, the sign of the error of the approximate SCGF at a given truncation of the cumulant expansion can change resulting in the cancellation of errors of two oppositely signed cumulant corrections and an accidental near agreement of the exact SCGF. We have found that by considering the convergence of the consecutive terms of the cumulant expansion we can reliably determine the accuracy of the approximate SCGF.

\begin{figure*}[t]
\includegraphics[scale=0.3]{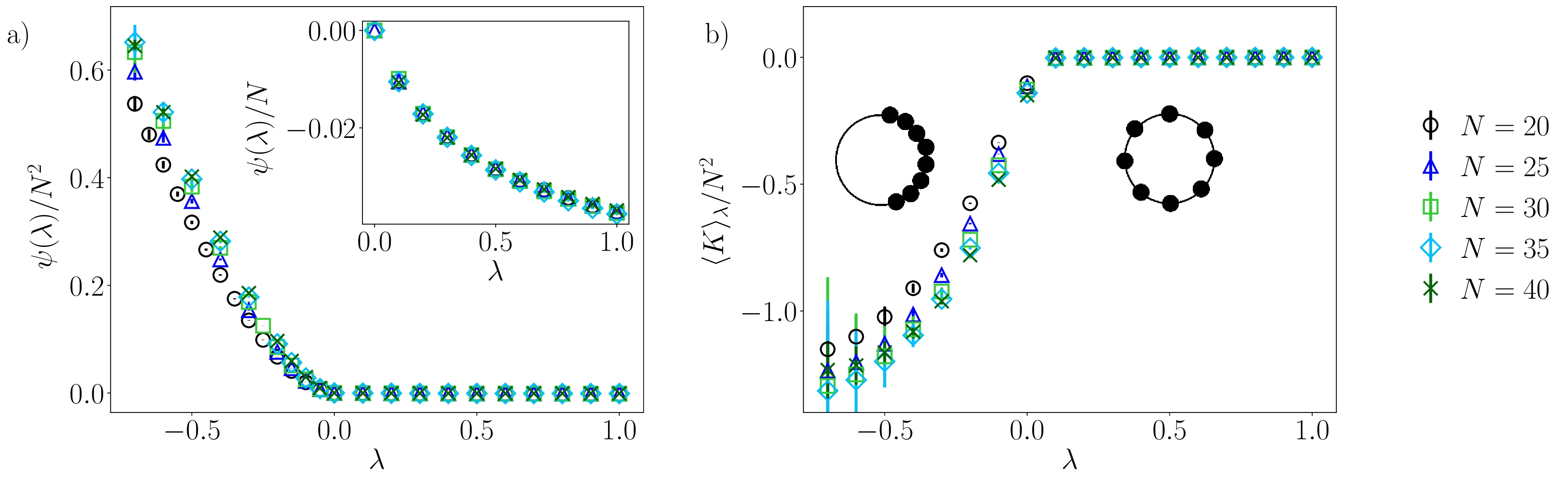}
\caption{Size-scaling of activity fluctuations of repulsive particles on a line. a) \(O(N^{2})\) scaling of \(\psi(\lambda)\) in the phase-separated state. (\textit{Inset}) \(O(N)\) scaling in the hyperuniform state.  b) Change in mean activity across the dynamical phase transition. (\textit{Inset}) Schematic representation of the phase-separated (left) and hyperuniform (right) states.}
\label{Fig4}
\end{figure*}

\subsection{Activity fluctuations of overdamped repulsive particles}

To study how this algorithm performs in an interacting system, we consider the fluctuations of the activity in a system of overdamped repulsive particles on a line. 
In both lattice and continuum models of volume excluding particles in one dimension, it has been reported that there are two characteristically distinct types of activity fluctuations, with a dynamical phase transition separating them.\cite{jack2015hyperuniformity} For rare large negative values of the activity, such systems spontaneously phase separate into macroscopically sized clusters, whereas for rare small values of the activity, they form a hyperuniform phase in which long-wavelength density fluctuations are suppressed. This behavior emerges as a singularity in the SCGF and a closing of the gap in the eigenspectrum of the tilted operator, which in the hydrodynamic scaling limit is predicted to occur with a critical point at \(\lambda_{c}\to0^{-}\).\cite{jack2015hyperuniformity,dolezal2019large} This system is thus suitable to test the effectiveness of the variational algorithm in computing rare fluctuations that are collective in origin.

Specifically, we study the fluctuations of dynamical activity in a system of \(N\) overdamped repulsive particles in a one-dimensional periodic box of length \(L\). The equation of motion is
\begin{equation}
\gamma\dot{x}_{i}=F_{i}(\mathbf{x})+\eta_{i}
\end{equation}
where \(F_{i}(\mathbf{x})\) is the total force felt by the \(i\)-th particle, 
\begin{align}
F_{i}(\mathbf{x})&=-\frac{\partial}{\partial x_{i}}\sum_{j\neq i}V_\mathrm{WCA}(x_{ij})
\end{align}
where $x_{ij} = x_{i}-x_{j}$ and the force is derived from a WCA pair potential
\begin{align}
V_\mathrm{WCA}(r)&= \left[ 4\epsilon\left( \frac{\sigma^{12}}{r^{12}}-\frac{\sigma^{6}}{r^{6}} \right) +\epsilon\right] \, , \quad r<2^{1/6}\sigma \\
&=0 \, , \quad  r\geq2^{1/6}\sigma\nonumber
\end{align}
with characteristic energy, \(\epsilon\), and length scale, \(\sigma\). The Gaussian white noise,  \(\eta_{i}\), is specified by
\begin{equation}
\langle\eta_{i}(t)\rangle=0 \, , \hspace{1cm}\langle\eta_{i}(t)\eta_{j}(t')\rangle=2 \gamma \kB T\delta_{ij} \delta(t-t')
\end{equation}
We work with \(\kB T=0.5\), \(\gamma=1\) and \(\sigma=1\). As before, we define our unit of time for this system as \(2\kB T/\gamma\sigma^{2}\) and we have reported all observables in dimensionless units. Additionally, we set \(\epsilon=1\) and consider a density of \(\rho=N\sigma/L=0.5\), so that the box is half-filled.

We study a measure of activity derived from the probability that the particles stay in the same state in a short time interval.\cite{autieri2009dominant} This form of the activity,
\begin{equation}\label{eq:act}
K_{\tau}=\frac{1}{\tau}\int_{0}^{\tau}dt\sum_{i}\left( \frac{F_{i}^{2}}{4\gamma \kB T}+\frac{1}{2\gamma}\frac{\partial F_{i}}{\partial x_{i}} \right)
\end{equation}
is also a part of the time-symmetric component of the path-action,\cite{maes2019frenesy} and its long time statistics are similar to other commonly used metrics that count the total number of hops for particles on a lattice.\cite{pitard2011dynamic, fullerton2013dynamical}
Using Ito's Lemma to simplify the last term in Eq. (\ref{eq:cmlt2}), the variational expression for the SCGF becomes
\begin{align}
\psi(\lambda)=\sup_{\mathbf{u}(x_{1},x_{2},...,x_{N})}
\frac{1}{\tau}\biggl\langle \int_{0}^{\tau}dt\sum_{i}\biggl[ \lambda\biggl( \frac{F_{i}^{2}}{4\gamma \kB T}+\frac{1}{2\gamma}\frac{\partial F_{i}}{\partial x_{i}} &\biggr) \nonumber\\
+\frac{u_{i}^{2}-F_{i}^{2}}{4\gamma \kB T}+\frac{1}{2\gamma}\frac{\partial (u_{i}-F_{i})}{\partial x_{i}} \biggr] &\biggr\rangle _{\mathbf{u}}
\end{align}
where in addition to the force, we require the gradient of both the original and the control force. 

For this system, the optimal control force \(\mathbf{u}(\mathbf{x})\) is in general long-range and many-bodied. 
Previous work on related one-dimensional systems have shown long-range repulsive interactions stabilizing the hyperuniform state for values of activity small in magnitude,\cite{simon2010asymmetric} and long-range attractive forces acting on the surface of particle clusters that emerge in rare large negative fluctuations of the activity.\cite{dolezal2019large} For our variational ansatz, we have approximated the many-body force as a sum of long-range pairwise interactions. Pair forces are the lowest rank approximation to this system due to its translational invariance. From the Hopf-Cole transform, optimization of a pair force is analogous to optimization of a two-body Jastrow function as used in variational quantum Monte Carlo.\cite{neuscamman2013jastrow} 

To represent the control force, we expand it in a basis of Laguerre polynomials \(L_{p}\) with coefficients \(c_{p}\) as
\begin{align}\label{eq:u2}
\tilde{u}_{i}=\sum_{j\neq i}\biggl[ -\frac{\partial}{\partial x_{i}}V_\mathrm{WCA}(x_{ij})+\sum_{p=1}^{M_{3}}c_{p}\left( L_{p}(\tilde{x}_{ij})e^{-\tilde{x}_{ij}/2}-1\right) \frac{x_{ij}}{|  x_{ij} |}\biggr]
\end{align}
where $\tilde{x}_{ij}= \alpha - \beta |x_{ij}|$ is a linear transformation on the distance between particles $i$ and $j$. The parameters $\alpha$ and $\beta$  can be adjusted to set a scale and a cutoff for where the force smoothly decays to zero, and \(M_{3}\) determines the size of the basis. The basis is complete for all possible two-body forces in the limit of large \(M_{3}\). The exponential factor makes the basis functions orthogonal and aids in the convergence of the optimization. We have used \(M_{3}=10\) for all of our results. We have fixed \(\beta=2/L\), and optimized \(\{c_{p}\}\) and \(\alpha\) with starting values of 0 and \(L/2\) respectively. In each iteration of the optimization, a trajectory of length \(2\times 10^{4}\) time units is simulated, the first half again reserved for equilibration and the second half being used to compute the gradients. For computing the integrated correlation function in Eq. (\ref{eq:tav2}), we have used \(\Delta t=200\) units.  
After obtaining the optimized control force in this ansatz, we use it to compute the unbiased SCGF using a cumulant expansion as before, with an observation time \(\tau=10\) and a total trajectory length of \(5\times10^{4}\) units. Across the range of $\lambda$ considered, we find convergence using the first three cumulants to correct the variational result. The SCGF obtained from this cumulant expansion is identical to results obtained using a guided cloning algorithm that has been described later in this section.

\begin{figure}[b]
\includegraphics[scale=0.3]{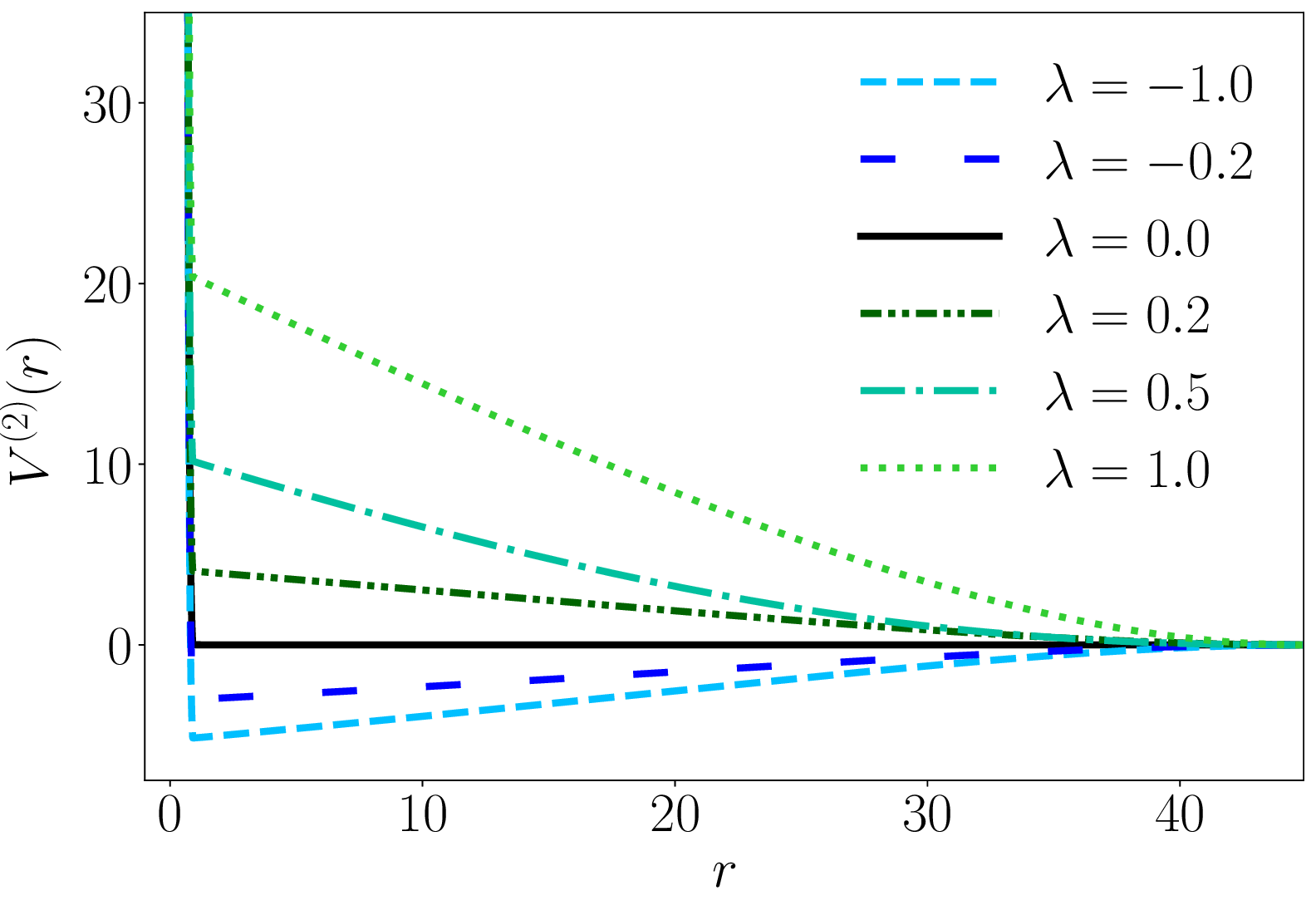}
\caption{Optimal pair-potential for positive and negative \(\lambda\) for \(N=40\).}
\label{Fig5}
\end{figure}

In Figure~\ref{Fig4}(a) we have plotted the size scaled SCGF, and the mean activity, for positive and negative values of \(\lambda\). For \(\lambda>0\), we find the system in a hyperuniform state, where all particles are pushed apart from each other and long-range density fluctuations are suppressed.\cite{jack2015hyperuniformity} The SCGF is size-extensive in this range of $\lambda$. For \(\lambda\ll0\) the particles phase separate, forming a single cluster. In the region where \(\lambda\) is negative but small, there is a phase transition to this clustered state accompanied by an inflection point in the mean activity, shown in Fig.~\ref{Fig4}(b), obtained from taking the numerical derivative of the SCGF, \(\langle K\rangle_{\lambda}=\psi'(\lambda)\). The extensive scaling regime has been explored systematically in a related model and found to agree well with predictions from macroscopical fluctuation theory.\cite{dolezal2019large} In our studies, we find it limited to $0>\lambda > -0.02$.
For large negative values of $\lambda$,  the cluster is a highly compressed solid with system-spanning correlations that result in the SCGF scaling super-extensively. In this regime of the SCGF, the typical force is on the order of $\sqrt{N}$, and can continue to increase with decreasing $\lambda$ because of the soft repulsion of the WCA potential. Inspection of the distribution of mean squared forces reveals that the cluster is not homogeneous, but most compressed in its interior with lower density near the edges, with a system size independent profile, see Appendix C. 
The phase transition from a disordered state to a clustered state is in accord with previous observations in related systems, and result in diverging correlation times rendering the precise study of the critical point difficult.\cite{dolezal2019large,jack2015hyperuniformity} We therefore focus our attention on the two phases on either side of that transition. Error bars were obtained from independent statistics from 3 distinct trajectories.

Figure~\ref{Fig5} shows the effective pair-potential, \(V^{(2)}(r)\), derived from the optimal control force at different values of \(\lambda\), for \(N=40\), obtained by the numerical integration of the control force. The potential is long-ranged and repulsive in the hyperuniform phase, and long-ranged and attractive in the clustered phase. The long-range potential leads to the observed size scaling in Fig.~\ref{Fig4}, because it imposes infinite range correlations. We also observe that the depth of the attractive potential for increasingly negative values of \(\lambda\) tends to saturate, while the magnitude of the repulsive potential for increasingly positive \(\lambda\) does not. This difference arises from the steeply rising WCA forces that can achieve more negative values of \(\langle K\rangle_{\lambda}\) with just a slight decrease in the nearest neighbor distance in the controlled system. In the hyperuniform phase, achieving the rarer values of activity implies an exponentially small number of collisions between the particles, which leads to an increasing repulsive control force. These optimal control forces derived from the variational ansatz do not contain many-body components unlike analytically derived approximate forces,\cite{dolezal2019large} yet they achieve the same phenomenology of phase separation and hyperuniformity described previously. 

\begin{figure}[t]
\includegraphics[scale=0.3]{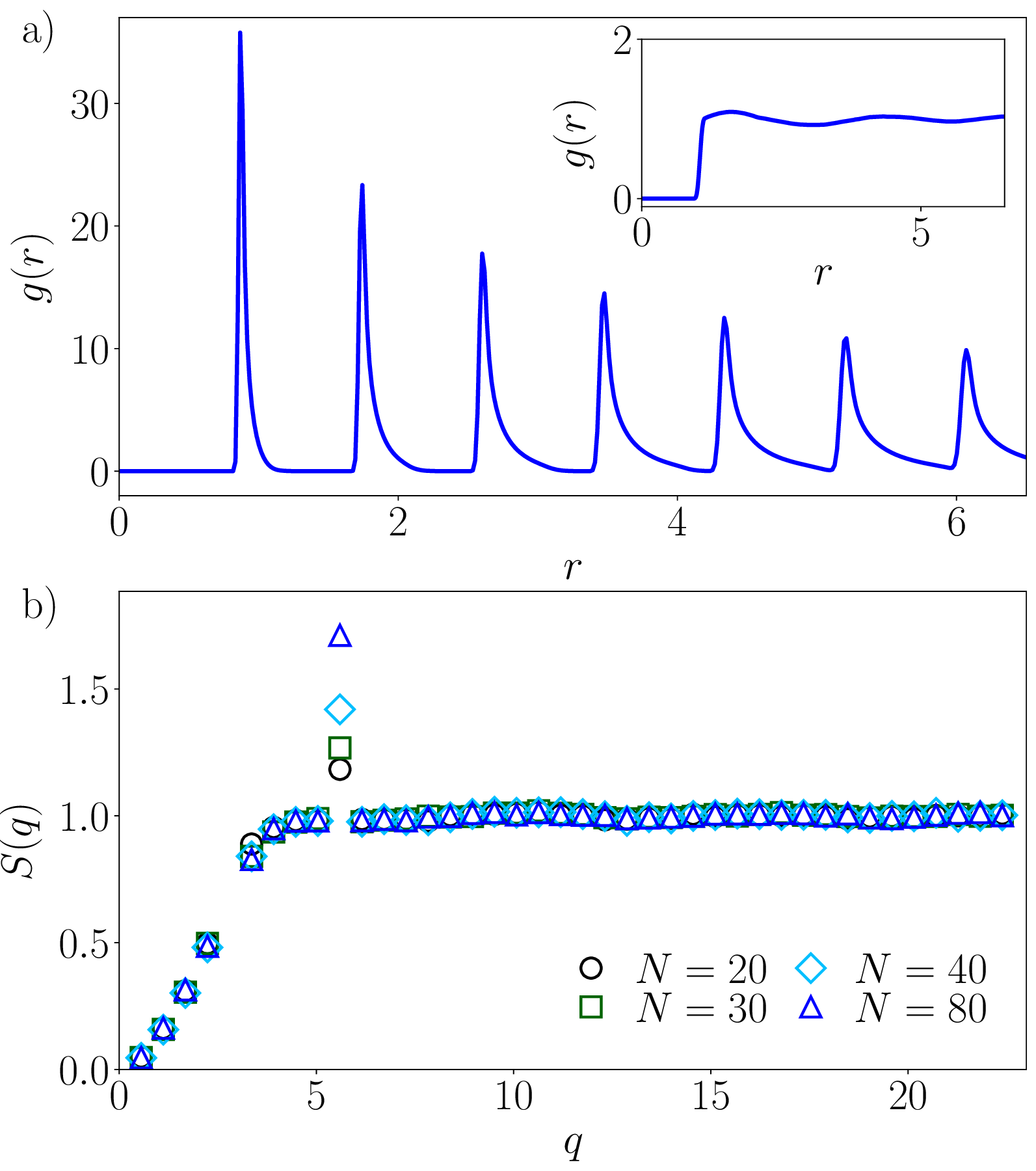}
\caption{Characterization of the two dynamical phases for \(N=80\). a) Pair distribution functions within the phase separated, \(\lambda=-0.1\), and (\textit{inset}) hyperuniform, \(\lambda=0.1\), states. b) Structure factor for various system sizes in the hyperuniform state, \(\lambda=1\).}
\label{Fig6}
\end{figure}

Figure ~\ref{Fig6}(a) characterizes the steady-state radial distribution function \(g(r)\),
\begin{equation}
\rho g(r) = N \langle  \delta(r-|x_{12}|) \rangle_{{\mathbf{u}}}
\end{equation}
 obtained in these phases, for a system size of \(N=80\), where \(x_{12}\) denotes the interparticle distance between each distinct pair of particles. In the phase-separated state, the particles form a solid cluster that has sharp peaks in \(g(r)\) at  intervals of $\sigma$. In the hyperuniform phase, the particles are repelled away from each other and \(g(r)\) has little structure aside from the volume-exclusion. We also characterize the structure of the hyperuniform state through the structure factor, \(S(q)\), as a function of the wavenumber \(q\), obtained from
\begin{equation}
S(q)=\frac{1}{L}\left \langle \left | \sum_{j=1}^{N}e^{-iqx_{j}} \right |^{2}\right \rangle_{\mathbf{u}}
\end{equation}
where the averages are computed in the ensemble with the control force. A linear increase of $S(q)$ from zero at small \(q\) is a signature of the suppression of long-wavelength density fluctuations in the hyperuniform phase, which we confirm in Fig.~\ref{Fig6}(b). The spike at $q=2\pi/2^{1/6}\sigma$ results from \(2^{1/6}\sigma\) being the distance of closest approach of the repulsive particles without experiencing a force.

\begin{figure}[b]
\includegraphics[scale=0.3]{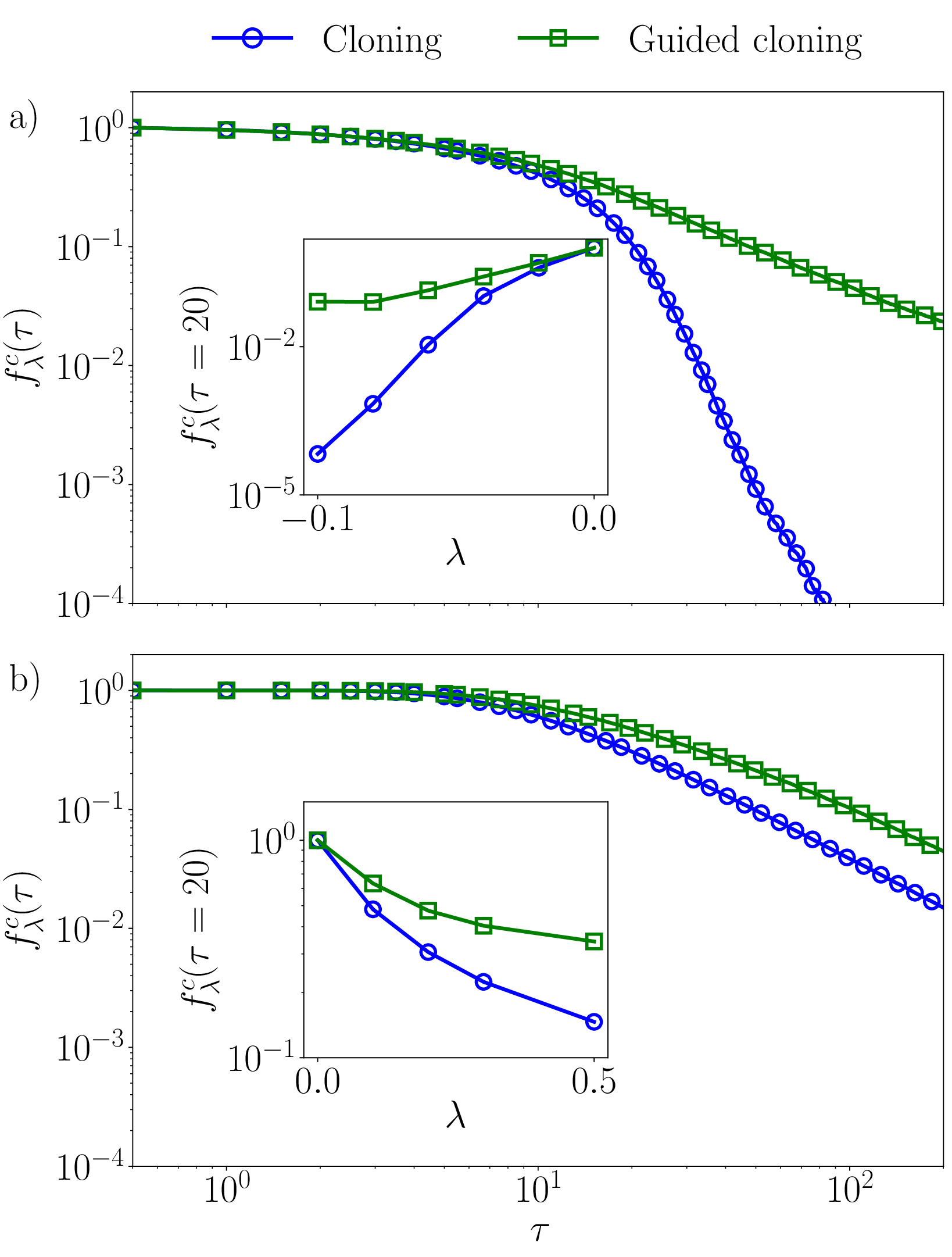}
\caption{Improvement of walker statistics of the cloning algorithm using approximate control forces as guiding functions in an N=20 system, represented by \(f_{\lambda}^{c}(\tau)=N_{c}(\tau)/N_{w}(\tau)\),  after an observation time \(\tau\). Blue circles are without a guiding force and green squares are with the variationally optimized guiding force. Decay of the fraction of uncorrelated walkers with increasing observation time in a) the phase-separated state (\(\lambda=-0.04\)) and b) the hyperuniform state \((\lambda=0.2\)). (\textit{Insets}) Decay of the fraction of uncorrelated walkers after \(\tau=20\) as a function of \(\lambda\) in a) the phase-separated state (\(\lambda=-0.04\)) and b) the hyperuniform state \((\lambda=0.2\)).}
\label{Fig7}
\end{figure}

While we have not investigated the phase transition directly, the disparate behavior of either side of the dynamical phase transition provides a useful test of our ability to obtain control forces, as the structure and dynamics of the system in the phase separated and hyperuniform states are very different. Despite their differences in both regimes, we are able to obtain control forces that are near enough to the optimal force to converge the large deviation functions using a brute force evaluation of the remaining cumulant expansion. Nevertheless, we expect this strategy may fail in general, in which case a more robust means of estimating the remaining contribution must be employed. To explore such alternatives, we apply these control forces as guiding functions within the cloning algorithm.\cite{ray2018exact}
To quantify the statistical benefit from the control forces, we start with a trajectory ensemble of \(N_w=32000\) walkers and monitor the decay rate in the number of uncorrelated walkers, $N_c$, with and without the control forces. 
The number of uncorrelated walkers is defined as those with a distinct history, having not been previously merged into an existing walker.
Figures ~\ref{Fig7}(a) and (b) show the statistics of the walkers with respect to observation time, with and without the control forces, in a system with 20 particles, and branching steps taken every 0.5 time units. We have plotted \(f^{c}_{\lambda}(\tau)=N_{c}(\tau)/N_{w}(\tau)\), where \(\tau\) is the observation time, to represent the growth of correlation in the trajectory ensemble.

In the clustered state, incorporating the control forces improves the number of uncorrelated walkers by multiple orders of magnitude. For larger negative \(\lambda\), an unbiased estimate of the SCGF can be obtained only when the variational control forces are used. The improvement in the statistics of the walkers increases for more negative \(\lambda\) because the magnitude of the SCGF grows rapidly, and therefore the weight carried by the branching step increases. We see this effect in the inset, where we show the fraction of uncorrelated walkers left after an observation time and how it varies with \(\lambda\).\cite{ray2018exact}

The decay of the walkers depends on the overlap between the tilted trajectory ensemble and that generated from the controlled dynamics. Slower decay will result when the control dynamics generates a trajectory ensemble that is close, in this sense, to the tilted trajectory ensemble. This behavior is analogous to other approximate guiding function based importance sampling, such as that arrived by iterative feedback\cite{nemoto2016population} or analytical approximation.\cite{dolezal2019large}
These effects are seen in the hyperuniform phase as well, albeit the decay of walkers in the ordinary cloning algorithm is less drastic, and so is the improvement by incorporating the guiding forces. The improvement in statistical efficiency upon including the optimized forces is not restricted to the cloning algorithm, and could be analogously adopted within transition path sampling\cite{dolezal2019large} or forward flux sampling.\cite{allen2009forward}

\section{CONCLUSION}
We have developed a variational algorithm to compute optimal control forces for Langevin models driven into nonequilibrium steady-states. We have used the control forces to sample rare fluctuations in time integrated dynamical observables like current and activity, in order to compute large deviation functions, and shown that they can be used to improve the efficiency of the cloning algorithm. Our variational algorithm, along with the correction of the systematic error with the cumulant expansion, has improved scaling properties compared to trajectory ensemble methods, and can be useful in dealing with many-particle chemical or biological systems.

Though we worked with Langevin models of structureless particles, the algorithm is straightforward to generalize to higher dimensions, where optimal control forces might have significant rotational components. It can also be extended to lattice models, where the rate matrix has to be expressed in a variational ansatz. A system modeled by a different stochastic equation of motion, like that employing an Andersen thermostat\cite{frenkel2001understanding} or quantum trajectory-based approaches,\cite{segal2010numerically,schile2018studying} can also be treated through this algorithm by changing only the functional forms of the path-actions provided a Doob transformation exists. 

The versatility of the variational algorithm allows for its use with different force ansatzes. In the activity-biased system, using a low-rank approximation for a many-body optimal control force was sufficiently accurate. However in cases where the control force is not expressible in a simple functional form or even as a many-body expansion, machine learning using artificial neural networks could be used to approximate it. The variational algorithm relies on evaluating functional derivatives of the force with respect to the parameters, which can be automated with autodifferentiation algorithms,\cite{baydin2018automatic} as has already been demonstrated in equilibrium free energy calculations.\cite{bonati2019neural} The use of techniques developed in this paper can aid the formulation of such optimization algorithms in the future. Additionally, this algorithm can be used for model reduction in high-dimensional systems,\cite{hartmann2016model} and hence to extend Variational Force-Matching and Ultra Coarse Graining algorithms\cite{dama2013theory,davtyan2014theory,dama2017theory} out of equilibrium, so that biomolecular and other soft matter systems can be simulated over large length and time scales with effective forces in nonequilibrium steady-states.

Lastly, this framework of solving the optimal forces can tackle  inverse-design problems out of equilibrium. Various inverse-design algorithms have been proposed that can obtain optimal forces to rationalize materials design with targeted properties and to guide directed self-assembly of smaller objects.\cite{torquato2009inverse,miskin2016turning} Our variational algorithm can be used to obtain optimal forces suitable for targeted assembly or tailored particle distributions when nonequilibrium driving forces are present, and hence can be used to characterize and predict dynamical phases in new functional materials.

\begin{acknowledgements}
We thank Robert L. Jack and Hugo Touchette for helpful discussions. This work was supported by the UC Berkeley College of Chemistry. We also acknowledge support from the U.S. Department of Energy,  Office of Basic Energy Sciences through Award Number DE-SC0019375.
\end{acknowledgements}

\section*{APPENDIX}
\subsection{Choice of \(\Delta t\) for Malliavin weights}

\begin{figure}[t]
\includegraphics[scale=0.3]{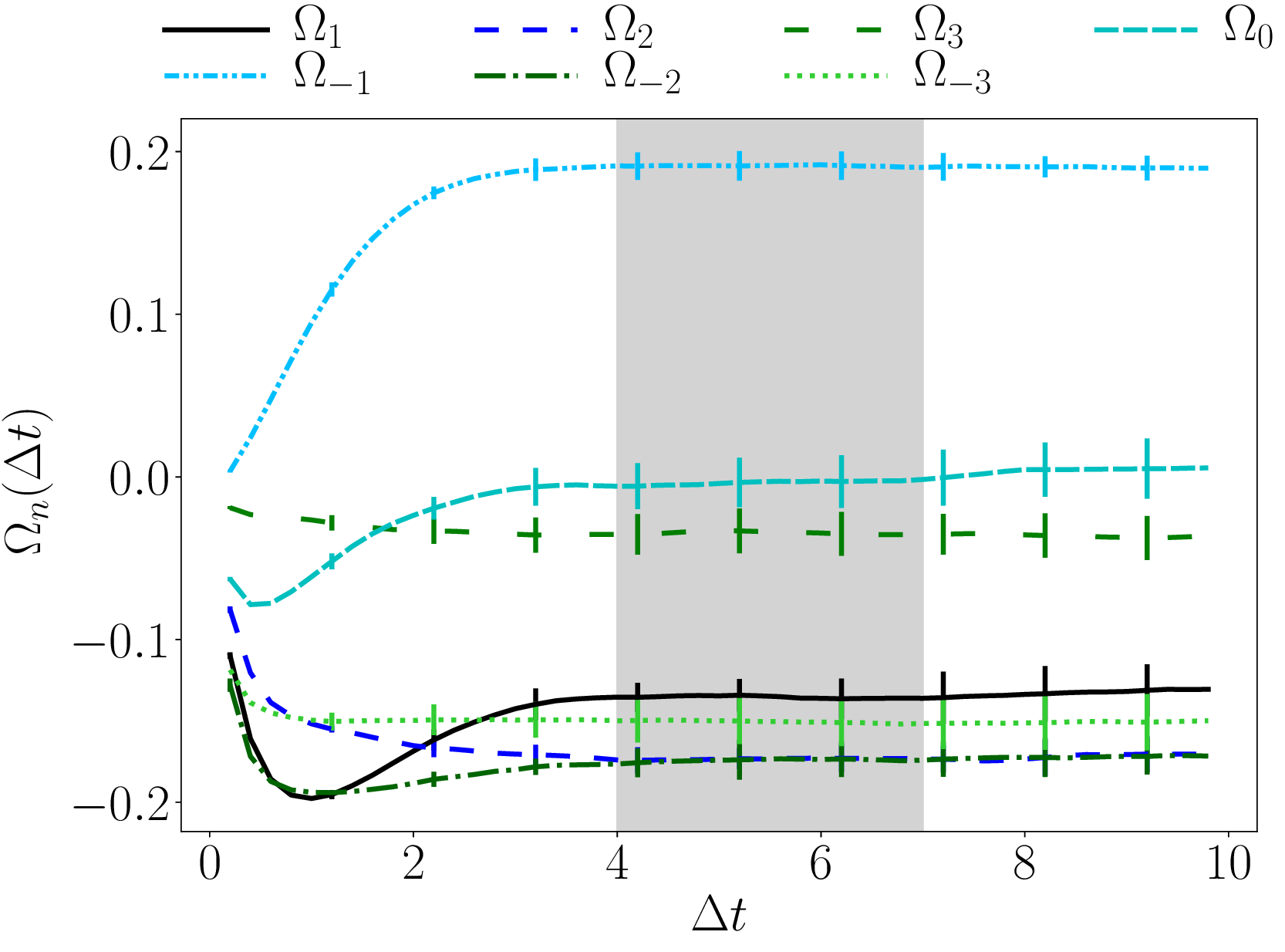}
\caption{Convergence of \(\Omega_{n}(\Delta t)\) for \(\lambda=0.5\), for different \(n\). Shaded region represents optimal choice of \(\Delta t\) for gradient descent.}
\label{Fig8}
\end{figure}

The choice of a finite integration limit \(\Delta t\) to compute the integral in Eq. (\ref{eq:tav1}) depends on both the intrinsic correlation times of the system and the timescale of the variance of the integrated correlation function to diverge. To illustrate this, we plot
\begin{align}
\Omega_{n}(\Delta t)&=\int_0^{\Delta t} dt \, \left\langle \delta \dot{y}_{n}(0) \delta\dot{O}[\mathbf{u}](t)\right\rangle _{\mathbf{u}}
\end{align}
for the system in Section IIIA in the \(m/\gamma t^{*}\to0\) limit. The ansatz can be written in this limit as
\begin{equation}
\tilde{u}(x)=F(x)+c_{0}+\sum_{n=1}^{3}[c_{n}\cos x+c_{-n}\sin x]
\end{equation}
and for Fig.~\ref{Fig8}, we have chosen \(c_{n}\) parameter values randomly between \(-1\) and \(1\), with \(\lambda=0.5\). We see that even though the correlation function converges for large \(\Delta t\), the error in the computed gradient increases steadily. For all the results in this paper, \(\Delta t\) was chosen to balance between these two effects so that the computed gradients suffers from no systematic error and minimum statistical error.

\subsection{Convergence of gradient descent}

\begin{figure}[b]
\includegraphics[scale=0.3]{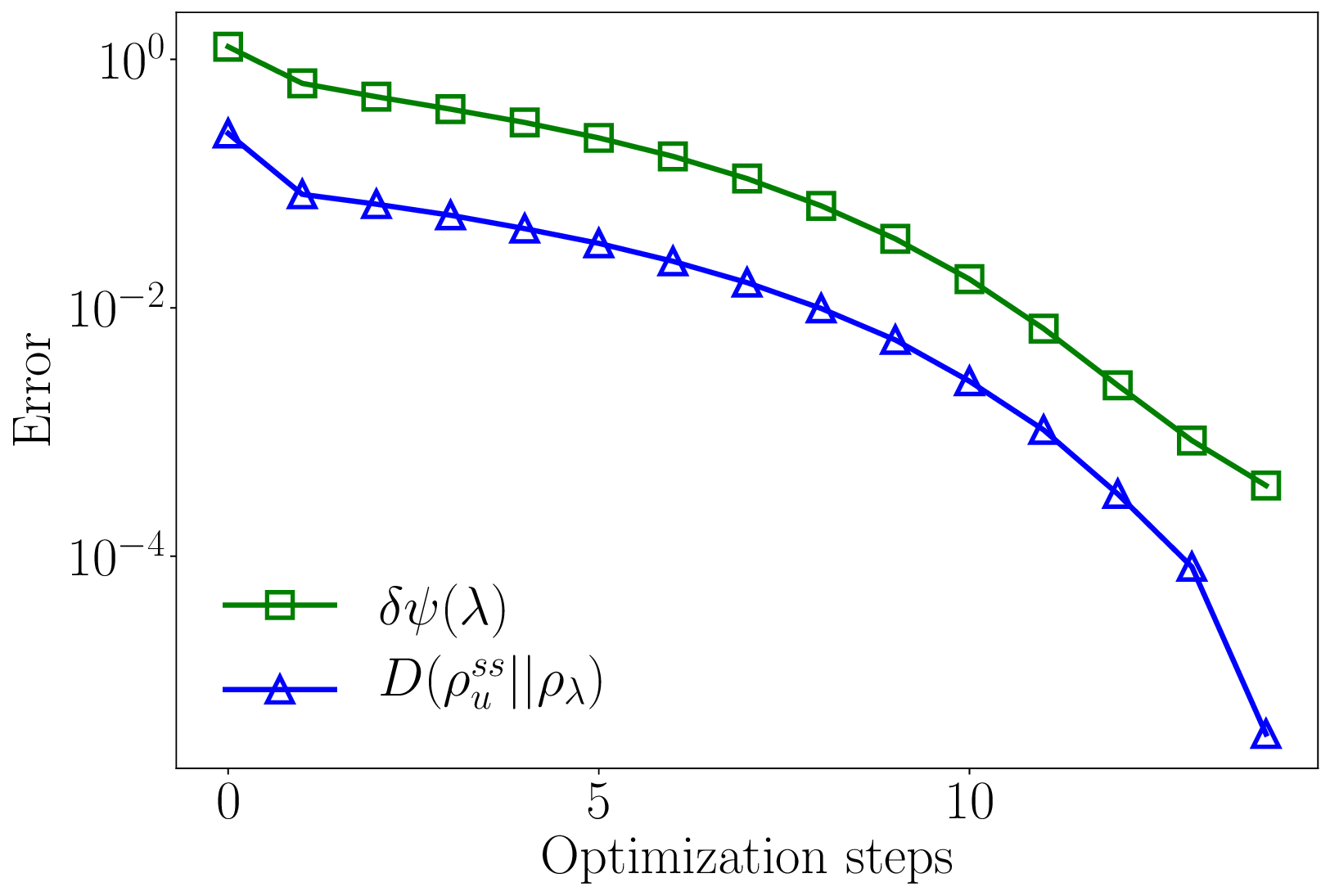}
\caption{Simultaneous convergence of SCGF and biased density for \(\lambda=0.5\).}
\label{Fig9}
\end{figure}

The accelerated gradient descent algorithm converges superlinearly, and in Fig.~\ref{Fig9} we have plotted the decrease of the systematic error \(\delta\psi(\lambda)\) in the current SCGF estimate with optimization steps, for the model system in Section IIIA, in the limit \(m/\gamma t^{*}\to0\). We also show the simultaneous convergence of the controlled ensemble steady-state density \(\rho^{ss}_{u}(x)\) to the true biased steady-state density \(\rho_{\lambda}(x)\propto\chi_{\lambda}(x)\phi_{\lambda}(x)\) where \(\chi_{\lambda}\) and \(\phi_{\lambda}\) are the dominant left and right eigenvectors of the tilted generator (\ref{eq:currentltilt}). We demonstrate this by plotting the relative entropy of the two,
\begin{equation}
D(\rho^{ss}_{u}||\rho_{\lambda})=\int dx\, \rho^{ss}_{u}(x)\log\left( \frac{\rho^{ss}_{u}(x)}{\rho_{\lambda}(x)} \right)
\end{equation}
which shows that even as only the current is being optimized to have a nontypical value, the entire trajectory ensemble simultaneously converges to the exact biased ensemble.

\subsection{Activity profile in clustered state}

\begin{figure}[b]
\includegraphics[scale=0.3]{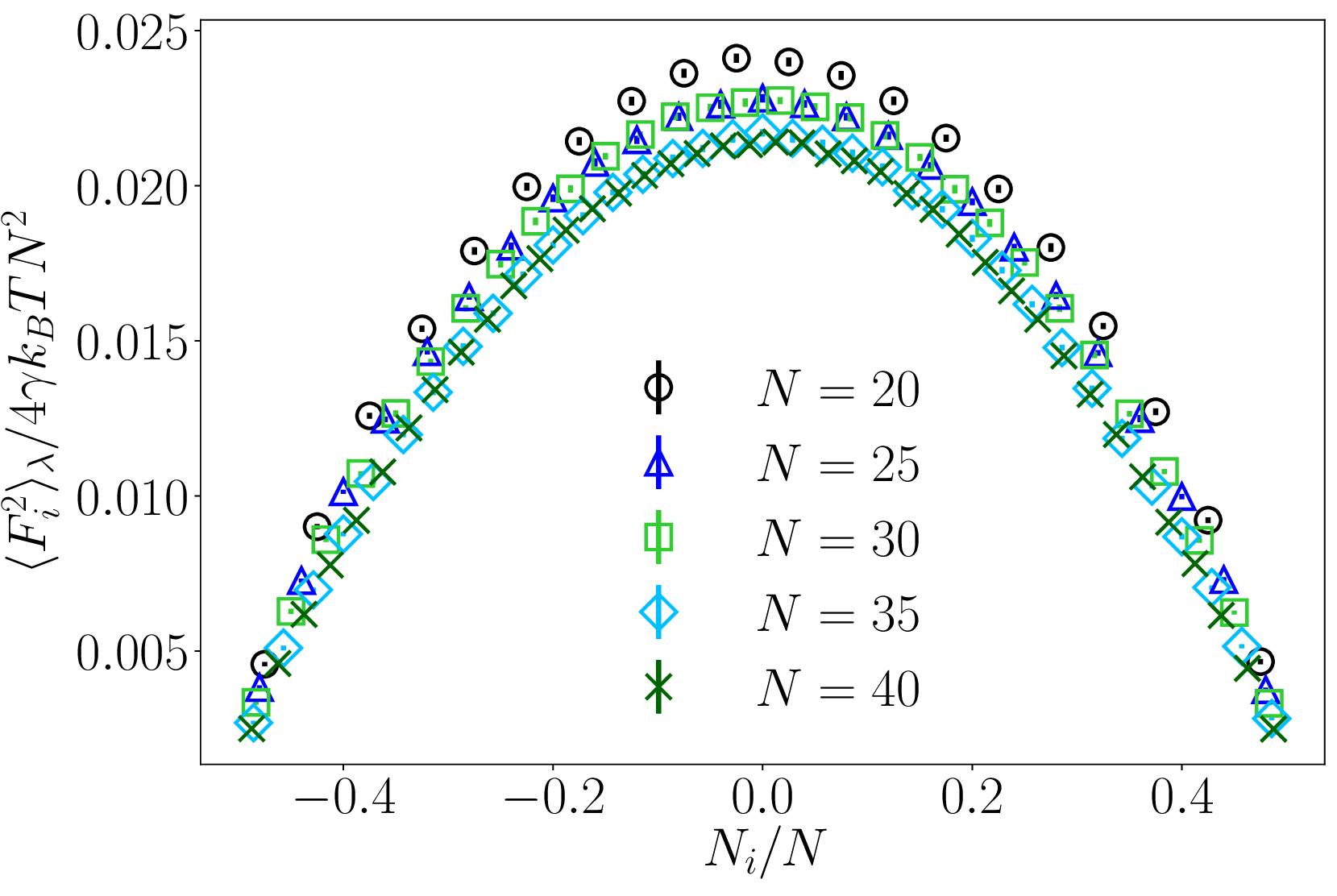}
\caption{Size-scaling of the mean squared force profile within the cluster for \(\lambda=-0.1\).}
\label{Fig10}
\end{figure}

Under large negative activity bias, we find that the overdamped repulsive particles form a highly compressed cluster. This cluster is described by system-spanning correlations. Shown in Fig.~\ref{Fig10} is the size-scaled profile for the first term of the collective activity (\ref{eq:act}), \(\langle F_{i}^{2}\rangle_{\lambda}/4\gamma\kB T\), with respect to a size-scaled particle index \(N_{i}=i-(N+1)/2\). The particles are indexed from one end of the cluster to the other, such that the center of the cluster is indexed at \(N_i=0\). The compressed cluster does not break apart during the duration of the trajectories observed, so that large \(|N_i|\) unambiguously refers to particles close to the surface of the cluster. The total mean activity \(\langle K\rangle_{\lambda}\) is proportional to the total mean squared force appearing in the first term, such that the profile of the second term in the definition looks analogous only with an opposite sign.\cite{fullerton2013dynamical} The \(O(N^{2})\) scaling of the mean squared force and its size-invariant parabolic profile explains the super-extensive SCGF scaling and the system spanning correlations in this \(\lambda\) regime.

\section*{REFERENCES}
\end{document}